\documentclass[conference]{IEEEtran}

\usepackage{amsmath}
\usepackage{amsfonts}
\usepackage{amssymb}
\usepackage{amsmath,amsthm}
\usepackage{color}
\usepackage{lmodern}
\usepackage{pgfplots}
\usepackage{algorithm}
\usepackage[noend]{algpseudocode}
\usepackage{graphicx}
\usepackage[export]{adjustbox}

\newtheorem{mydef}{Definition}
\newtheorem{myprop}{Proposition}

\begin{document}

\title{Improving the coding speed of erasure codes with polynomial ring transforms}

\author{
\IEEEauthorblockN{Jonathan Detchart,     J\'er\^ome Lacan}
\IEEEauthorblockA{\\Universit\'e de Toulouse ISAE-SUPAERO, Toulouse, France\\\{jonathan.detchart, jerome.lacan\}@isae-supaero.fr}
} 

\maketitle

\begin{abstract}
Erasure codes are widely used in today's storage systems to cope with failures. Most of them use the finite field arithmetic. In this paper, we propose an implementation and a coding speed evaluation of an original method called PYRIT (PolYnomial RIng Transform) to perform operations between elements of a finite field into a bigger ring by using fast transforms between these two structures. Working in such a ring is much easier than working in a finite field. Firstly, it reduces the coding complexity by design. Secondly, it allows simple but efficient \texttt{xor}-based implementations by unrolling the operations thanks to the properties of the ring structure. We evaluate this proposition for Maximum Distance Separable erasure codes and we show that our method has better performances than common codes. Compared to the best known implementations, the coding speeds are increased by a factor varying from $1.5$ to $2$.
\end{abstract}

\section{Introduction}
\label{sec:intro}
In today's storage systems, erasure codes are widely used and provide reliability to failures. They are used by RAID solutions \cite{Anvin:2009}, Cloud storage \cite{Huang:2012:ECW:2342821.2342823}, or as an elementary building block in large scale coding systems \cite{xorbas}. They replace replication by reducing the amount of extra storage needed to tolerate the same amount of erasures (\cite{Weatherspoon2002}). 
But this kind of technique is limited by the complexity of the arithmetic used. Most of the complexity of erasure codes consists in making linear combinations over a finite field. To speedup these linear combinations, several solutions have been presented:

Recently, the use of SIMD instructions have been proposed to drastically increase the speed of erasure codes, particularly by optimizing the multiplication of a large region of elements of a finite field by a constant element: \cite{Li:2008:PNC:1491269.1493312, 4770564, pgm:13:sfg}. In \cite{xor-luby}, the authors consider the elements of a finite field of characteristic 2 as a binary matrix where the entries represent a \texttt{xor} between two parts of data. In \cite{isit}, we extended an idea proposed in \cite{ITOH1989} by defining other transforms to speedup the coding process in the context of erasure codes. The contribution of this paper is essentially theoretical. 

In this paper, we propose an extension to other finite fields and a complete performance analysis of the method introduced in \cite{isit}, called PYRIT (PolYnomial RIng Transform), replacing the multiplication in a finite field by the multiplication in a ring by using transforms between particular finite fields and polynomial rings. We show that an element of a field can have several representations in a bigger ring, and that the choice of this element can have an impact on the performances.
We also show that using a ring to perform multiplications allows to reduce the complexity of the coding process. It also allows some optimizations in the implementation which are not possible when using a classic \texttt{xor} based implementation. 
We compare our implementation with other implementations and show that we are faster than the best known implementations for both single and multithreading.

\section{Correspondence between field and ring}
\label{sec:algebra}

\subsection{Algebraic context}

Let us recall some known properties about polynomial rings and fields:

Let $p(x)$ be an irreducible polynomial of degree $w$. The field  $\mathbb{F}_{2}[x]/(p(x))$ is denoted by $\mathbb{F}_{2^w}$. The polynomial $x^n+1$ is not irreducible, and thus $\mathbb{F}_{2}[x]/(x^n+1)$ is a ring. Let us denote it by $R_{2,n}$.

Let us now consider the factorization of $x^n+1$ into irreducible polynomials: $x^n+1=p_1^{u_1}(x)p_2^{u_2}(x)\ldots p_l^{u_l}(x)$.
If $n$ is odd, it can be shown that $u_1=u_2=\ldots=u_l=1$ (see \cite{poli1992error}). In this document, we only consider this case.

The proofs of the following propositions can be found in \cite{poli1992error} or \cite{MacWilliams19711}.

\begin{myprop}
	\label{prop:directSumDecomposition}
	The ring $R_{2,n}$ is equal to the direct sum of the principal ideals of  $\mathbb{F}_{2}[x]$ $A_i=( (x^n+1)/p_i(x) )$ for $i=1,\ldots,l$. 
	Each ideal contains an unique idempotent $\theta_i(x)$. 	
\end{myprop}

\begin{myprop}
	\label{prop:isomorphism}	
	For each $i=1,\ldots,l$, $A_i$ is isomorphic to the finite field $B_i=\mathbb{F}_{2}[x]/(p_i(x))$. The isomorphism is :
	\begin{equation}
	\phi_i  :  \begin{array}{ccc}
	B_i & \rightarrow & A_i \\
	b(x) & \rightarrow & \bar{b}(x)=b(x)\theta_i(x) \
	\end{array}
	\end{equation}  	
	and the inverse isomorphism is :
	\begin{equation}
	\phi_i^{-1}  :  \begin{array}{ccc}
	A_i & \rightarrow & B_i \\
	a(x) & \rightarrow & a(x)\textrm{ mod }p_i(x)\\
	\end{array}
	\end{equation}  	
\end{myprop}

We will use this kind of morphism to transform each finite field element from the source vectors and  the generator matrix of the erasure code into ring elements to perform xor operations and apply reverse transforms on the generated vectors.

\subsection{The sparse transform for generator matrices}
\label{sec:sparse}
One of the main challenges in the erasure code construction is the choice of the generator matrix. Since we here only consider systematic codes, the parity part of the matrix must verify two main properties:
\begin{itemize}
	\item it must be as sparse as possible. Indeed, the number of \texttt{xor} done on the data is defined by the number of ones in the binary form of the generator matrix,
	\item any square submatrix must be invertible to verify the MDS property.
\end{itemize} 

For the first point, we use an interesting property of the correspondences between the ring and the field. The number of ring elements is $2^n$, which is greater than the number of the elements of the field: $2^w$.  

In fact, the important point behind the use of a ring is that one of its ideals, $A_1$, is isomorphic to the field. A naive approach to multiply two elements of the field $u(x)$ and $v(x)$ would consist in sending them in $A_1$ by applying the isomorphism $\phi_1$ to obtain $\bar{u}(x)$ and $\bar{v}(x)$. The second step would consist in multiplying  $\bar{u}(x)$ and $\bar{v}(x)$ and apply the inverse isomorphism $\phi_1^{-1}$ on the result.

However, we can observe that the structure of the ring (which is decomposed as a direct sum of ideals) allows to consider that the operations on the ring can be decomposed into "parallel" and independent operations in each ideal. It follows that in the function from the field to the ring, we can add other ring elements which belong to the other ideals. These do not interfere with the operations in $A_1$ which are the ones important for the field. 

To be more precise, let us define a function $\xi$ from the field to the ring which is such that :
for any $u(x)$ in the field, $\xi(u(x))=\bar{u}(x)+\hat{u}(x)$, where $\bar{u}(x)=\phi_1(u(x)) \in A_1$ and $\hat{u}(x)$ is an element of the ring which does not have a component in $A_1$ (see Proposition \ref{prop:directSumDecomposition}).

Then, to multiply the field elements $u(x)$ and $v(x)$, we can compute the product $\xi(u(x)).\xi(v(x))$ which is equal to $\bar{u}(x).\bar{v}(x)+\hat{u}(x).\hat{v}(x)$. The application $\phi^{-1}$ to this result removes the part outside $A_1$ and then outputs $u(x).v(x)$ in the field. This means that, to perform computations with element $u(x)$, we can use any element of the form $ \bar{u}(x)+\hat{u}(x)$ in the ring, where $\bar{u}(x)=\phi_1(u(x))$ and $\hat{u}(x)$ is an element of the ring which does not have a component in $A_1$. The number of ring elements which do not have a component in $A_1$ is equal to $2^{n-w}$. 

The interest of this property is that the complexity of the multiplication in the ring is not the same for all the elements. Indeed, for the generator matrix in the binary form, the complexity depends on the number of ones. So this transfom allows to choose the element in the ring having the lowest weight among the ring elements corresponding to a given field element. We call this operation the \texttt{sparse transform}.

\subsection{Pyrit using AOP}
\label{sec:aop}
All One Polynomials (AOP) of degree $w$ are equal to $x^w+x^{w-1}+x^{w-2}+\ldots+x+1$. The AOP of degree $w$ is irreducible over $\mathbb{F}_{2}$  if and only if $w+1$ is a prime and $w$ generates $\mathbb{F}^*_{w+1}$, where $\mathbb{F}^*_{w+1}$ is the multiplicative group in $\mathbb{F}_{w+1}$ \cite{Wah1984}. The values $w+1$, such that $w$ is an irreducible AOP is the sequence A001122 in \cite{A001122}.

The use of AOPs for fast operations in finite fields was studied by \cite{ITOH1989} and then by \cite{Silverman1999} in the context of hardware implementations of large finite field operations. 

Irreducible AOP of degree $w$ appears in the factorization of $x^{w+1}+1$ which is equal to $( x^w+x^{w-1}+x^{w-2}+\ldots+x+1).(x+1)$. Some of the previous propositions can be specified for these polynomials. 

Let $p(x)$ be the AOP of degree $w$. Then:
\begin{itemize}
	\item  the ring $R_{2,w+1}$ is equal to the direct sum of the principal ideals of  $\mathbb{F}_{2}[x]$ $A_1$ generated by $x+1$  and $A_2$ generated by $p(x)$. 
	The idempotent of $A_1$  and $A_2$  are respectively $p(x)+1$ and $p(x)$.
	\item  The isomorphism between $B_1=\mathbb{F}_{2}[x]/(p(x))=\mathbb{F}_{2^w}$ and $A_1$ is equal to 
	\begin{equation}
	\phi_1  :  \begin{array}{ccc}
	B_1 & \rightarrow & A_1 \\
	b(x) & \rightarrow & \bar{b}(x)=b(x)(p(x)+1) \
	\end{array}
	\end{equation}  	
	and the inverse isomorphism is:
	\begin{equation}
	\phi_i^{-1}  :  \begin{array}{ccc}
	A_i & \rightarrow & B_i \\
	\bar{b}(x) & \rightarrow & \bar{b}(x)\textrm{ mod }p(x)\\
	\end{array}
	\end{equation} 
\end{itemize}

An interesting property of $A_1$ is that it is equal to the set of polynomials of even weight. Indeed, since an element of $A_1$ is a multiple of $x+1$ (modulo $x^{w+1}+1$), it contains an even number of monomials. And since the number of elements of $A_1$, which is equal to $2^w$, corresponds to the number of polynomials of even weight, $A_1$ is equal to the set of even weight polynomials.

For example, by considering the finite field $\mathbb{F}_{2^4}$ defined by the irreducible All-One Polynomial $p(x) = 1 + x + x^2 + x^3 + x^4$, we can define $R_{2,5} = \mathbb{F}_{2}[x]/(x^5 + 1)$ as the quotient ring of polynomials of the polynomial $\mathbb{F}_2[x]$ quotiented by the ideal generated by the polynomial $x^5 -1$.

The polynomial $x^5+1$ is the product of the irreducible polynomials $p_1(x)=x^4 + x^3 + x^2 + x + 1$ and $p_2(x)=x+1$. The ring $R_{2,5}$ is the direct sum of the ideal $A_1$ generated by the polynomial $(x^5 + 1) /p_1(x) = p_2(x)$ and the ideal $A_2$ generated by the $(x^5 + 1) / p_2(x) = p_1(x)$. In others words, any element $u(x)$ of $R_{2,5}$ can be written in a unique way as the sum of two components $u_1(x)+u_2(x)$, where $u_1(x)\in A_1$ and $u_2(x)\in A_2$.
	It can be verified that $A_1$ (resp. $A_2$) contains one and only one idempotent $\theta_1(x)=x^4 + x^3 + x^2 + x$ (resp. $\theta_2(x)=x^4 + x^3 + x^2 + x+1$). A construction of this idempotent is given in \cite[Chap. 8, Theorem 6]{MWSl77}.

Since $\mathbb{F}_q[x]/(p_i(x))$ is isomorphic to $B_i = \mathbb{F}_{q^{w_i}}$ where $p_i(x)$ is of degree $w_i$, $R_{2,5}$ is isomorphic to the following cartesian product $R_{2,5} \simeq B_1 \otimes B_2$ with $B_1=\mathbb{F}_2[x]/(p(x))=\mathbb{F}_{2^4}$ and $B_2=\mathbb{F}_2[x]/(x-1)=\mathbb{F}_2$. The image by the isomorphism of  $b(x)\in B_1$ into $A_1$  is $\bar{b}(x)=\phi_1(b(x))= b(x)\theta_1(x) $. On the other side, the image of the element  $\bar{b}(x)$ of $A_1$ by the inverse isomorphism is equal to $b(x)=\phi_i^{-1}(\bar{b}(x))=\bar{b}(x)\textrm{ mod }p_1(x)$. 

	In $R_{2,5}$, let us consider $a(x)=1+x^2$ and $b(x)=x+x^4$ and their respective matrix and vector representations:
	\begin{displaymath}
	a(x) \longrightarrow \includegraphics[valign=m]{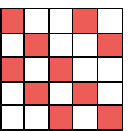}, \ b(x)  \longrightarrow    \includegraphics[valign=m]{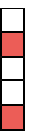}
	\end{displaymath}
	where the filled squares represent the ones and the empty squares represent the zeros.
	
	To compute the multiplication, we just have to perform the matrix vector multiplication:
	\begin{displaymath}
	a(x).b(x) \longrightarrow \includegraphics[valign=m]{figures/mat1x3.eps} * \includegraphics[valign=m]{figures/vectorxx4.eps} = \includegraphics[valign=m]{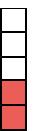} \longrightarrow  c(x)= x^3+x^4  
	\end{displaymath}
	We can verify that we obtain the same result with the multiplication of polynomials in the ring:
	\begin{eqnarray}
	a(x).b(x)& = & (1+x^2).(x+x^4)\textrm{ mod }(x^5+1) \nonumber\\
	& = & x^3+x^4 \nonumber
	\end{eqnarray}

	The field $\mathbb{F}_{2^4}=\mathbb{F}_{2}[x]/(x^4+x^3+x^2+x+1)$ is isomorphic to the ideal $A_1$ of the ring $R_{2,5}$ generated by $(x+1)$. Let us consider the element $u(x)=x^2$ of the field. According to Proposition \ref{prop:isomorphism}, its image $\bar{u}(x)$ in $A_1$ is equal to $u(x)\theta_1(x)=x^2.(x+x^2+x^3+x^4)=1+x+x^3+x^4$.
	According to the previous paragraph, to perform multiplications with this element in the ring, we can use any element of the form $\bar{u}(x)+r(x)$ where $r(x)$ does not have a component in $A_1$. The $2^{n-w}=2$ elements which have this property are $0$ and $p(x)=1+x+x^2+x^3+x^4$. So we can consider the binary matrices associated to $\bar{u}(x)$ and $\bar{u}(x)+p(x)$:
	\begin{displaymath}
	\bar{u}(x) \longrightarrow \includegraphics[valign=m]{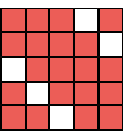} \ \ \   		
	\bar{u}(x)+p(x) \longrightarrow \includegraphics[valign=m]{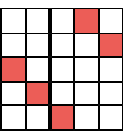}   
	\end{displaymath}
	We can observe that the matrix associated to  $\bar{u}(x)+p(x)$ is more sparse than the one associated to $\bar{u}(x)$. So this matrix is chosen to perform the multiplication corresponding to $u(w)$ in the ring.

The function which makes the correspondence between the elements of the field and the sparsest matrices among their corresponding ones is denoted by $\phi_S$. You can see an example of the binary representation for a generator matrix in figure 1. Note that in the example, the elements of the ring are represented by a 5x5 binary matrix. However, depending on the transforms applied to the data, some operations are useless, and we can remove the corresponding row or column. So the elements are represented by 4x5 or 5x4 matrices (see \ref{parity_transform} and \ref{emb_transform}).

\begin{figure}
\label{fig:gen_mat}
\begin{center}
\includegraphics[width=.8\linewidth]{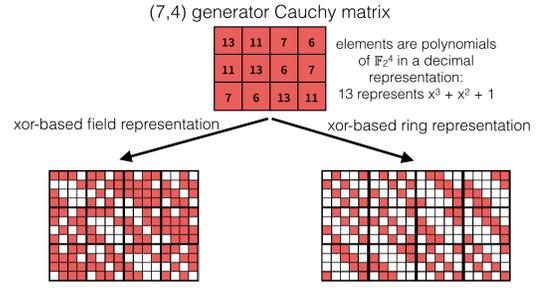}
\end{center}
\caption{binary generator matrices in a field and a ring}
\end{figure}

Now, we have to define the function that sends the data vector from the field into the ring. Even if the function  $\phi_S$ can be used, it is not optimal because the \texttt{xor}-based representation handles the data by blocks and not sequentially. Thus, it is not efficient to access the binary representation of each field element in order to determine the sparsest corresponding ring element. 

\subsubsection{The parity transform}
\label{parity_transform}
The first function we propose is to compute simple parity bits on the data blocks. In this case, the inverse operation just consists in removing the parity bits from the results of the matrix vector multiplication. As we explained in \cite{isit}, this transform is not an isomorphism, but just a bijection. That means operations in the field and in the ring are not compatible, and both encoder and decoder must use the parity transform to perform coding operations.

For an irreducible AOP of degree $w$, the usual isomorphism that sends the finite field elements into the ideal $A_1$ of the ring $R_{2,w+1}$ consists in multiplying the polynomial by the idempotent. However, we showed that $A_1$ is the set of polynomials of even weight. It follows that the function $\phi_P$ which adds a single parity bit to the vector corresponding to the finite field element (in order to have an even weight) can be used to make the correspondence between the field and the ideal. In the following equations, suppose $b(x) = a+b.x+c.x^2+d.x^3$ as an element of $\mathbb{F}_{2^4}$.

	\begin{displaymath}
		\phi_P : b(x) \rightarrow (a+b+c+d). x^4 + b(x)
	\end{displaymath}
For the inverse function, the vector resulting from the matrix vector multiplication contains elements of the ideal because the data vector belongs to the ideal and the matrix elements, obtained with $\phi_S$, belongs to the ring. So the finite field elements can be obtained by just removing the parity bit of each ring elements. We call this function $\phi_P^{-1}$. 

	\begin{displaymath}
		\phi_P^{-1} : e.x^4 + b(x) \rightarrow b(x)
	\end{displaymath}
To summarize, the $\phi_P$ function adds a single parity bit to the finite field elements and the inverse function removes it from the ring elements. These two functions can be very efficiently performed on the \texttt{xor}-based representation of the data. We can note that $\phi_P^{-1}$ is not an isomorphism. This implies that both the encoder and the decoder must use $\phi_P$ and $\phi_P^{-1}$.

\subsubsection{The embedding transform}
\label{emb_transform}

The second function we propose is a simple embedding, denoted by $\phi_E$ from the field in the ring. In other words, the polynomial corresponding to the finite field element is simply "padded" with $n-w$ zeros and considered as an element of the ring. 

	\begin{displaymath}
		\phi_E : b(x) \rightarrow b(x) + 0.x^4
	\end{displaymath}

For the inverse function, we use the the traditional inverse isomorphism $\phi^{-1}$ presented in Proposition \ref{prop:isomorphism} which corresponds to the computation modulo $p(x)$ where $p(x)$ is the irreducible polynomial. Note that these functions correspond to the ones proposed by \cite{ITOH1989}. According to the embedding function, the elements of the field are just padded by one $0$ and considered as elements of the ring. 

The inverse function is the computation of the remainder modulo $p(x)$. Since $p(x)$ is AOP, this operation consists in adding the last bit (the coefficient of the monomial of degree $w$) to all the other coefficients. 

	\begin{displaymath}
		\phi_E^{-1} : e.x^4 + b(x) \rightarrow (a+e)+(b+e).x+(c+e).x^2+(d+e).x^3
	\end{displaymath}

\subsection{Pyrit using ESP}
\label{sec:esp}

Similar approach can be used with $\mathbb{F}_{2^6}$. First, let's recall the ESP definition.
\begin{mydef}[\cite{ITOH1989}]
	
	\label{def:esp}	
	A polynomial $g(x)=x^{sr}+x^{s(r-1)}+x^{s(r-2)}+\ldots + x^s + 1=p(x^s)$, where $p(x)$ is an AOP of degree $r$, is called s-equally spaced polynomial ($s$-ESP) of degree $sr$.
	
\end{mydef}

According to \cite[Theorem 3]{ITOH1989}, $g(x)$ is irreducible if and only if $p(x)$ is irreducible and for some integer $t$, $s=(r+1)^{t-1}$ and $2^{r(r+1)^{t-2}}\not=1$ mod $(r+1)^t$.  

The first values of the pair $(r,s)$ for which the ESP is irreducible are $(2,3), (2,9), (4,5), (2,27),\ldots$ 

The ESP $g(x)=x^{sr}+x^{s(r-1)}+\ldots + x^s + 1$ divides the polynomials $x^{s.(r+1)}+ 1$ because $x^{s.(r+1)} + 1 = g(x).(x^s+1)$. Thus, according to Proposition \ref{prop:directSumDecomposition}, if the ESP $g(x)$ is irreducible, the field $\mathbb{F}_{2}[x]/(g(x))=\mathbb{F}_{2^{r.s}}$ is isomorphic to the ideal $A_1$ generated by $ (x^{s.(r+1)}-1)/g(x) = x^s+1$. 

It can be shown that the idempotent $\theta_1(x)$ of $A_1$ is equal to $g(x)+1$. 
Thus, the idempotent between the field $B=\mathbb{F}_{2}[x]/(g(x))=\mathbb{F}_{2^w}$ and the ideal $A_1$ are the following: 
\begin{equation}
\phi  :  \begin{array}{ccc}
B_1 & \rightarrow & A_1 \\
b(x) & \rightarrow & \bar{b}(x)=b(x)(g(x)+1) 
\end{array}
\end{equation}  	

and the inverse isomorphism is:
\begin{equation}
\phi^{-1}  :  \begin{array}{ccc}
A_1 & \rightarrow & B_1 \\
\bar{b}(x) & \rightarrow & \bar{b}(x)\textrm{ mod }g(x)\\
\end{array}
\end{equation} 

Like AOPs, ideals associated to ESP have an interesting parity property. Indeed, any element is a multiple of the generator polynomial $x^s+1$. So the element $a(x)$ is equal to $u(x).(x^s+1)=(\sum_{i=0}^{{sr}-1}u_i x^i).(x^s+1)$. $u(x)$ can also be expressed under the form $\sum_{j=0}^{s-1}x^jv_j(x^s)$ where $v_j(x)$ is a polynomial of degree $r-1$, for $j=0,\ldots,s-1$. Thus, we have  $u(x).(x^s+1)= \sum_{j=0}^{s-1}x^jv_j(x^s).(x^s+1)=\sum_{j=0}^{s-1}x^j v'_j(x^s)$ where $v'_j(x)=v_j(x).(x+1)$, for $j=0,\ldots,s-1$. Like for the AOP, this implies that the weight of each $v'_j$ is even. This means that any element of $A_1$ can be seen as the interleaving of $s$ even weight elements of length $r$. The number of elements which verify this property is exactly the number of elements of $A_1$, so this property characterizes the elements of $A_1$. 

In our case, we have $s=3$ and $r=2$. The ESP $g(x)=x^6+x^3+1$ is irreducible and allows to represent the finite field  $\mathbb{F}_{2^6}$. Its elements are sent onto the ring $R_{2,9}= \mathbb{F}_{2}[x]/(x^9+1)$ to perform fast operations.

\subsubsection{The parity transform}

As we said, the ideal corresponding to a finite field determined by an ESP is the set of elements which can be seen as an interleaving of $s$ even weight words of length $r$. Like AOP, we propose to add a single parity bit to each "interleaved" word of length $r$ in order to verify the parities. In the following equations, let's suppose $b(x) = a+b.x+c.x^2+d.x^3+e.x^4+f.x^5$ as an element of $\mathbb{F}_{2^6}$.

	\begin{displaymath}
		\phi_P : b(x) \rightarrow b(x) + (a+d).x^6+(b+e).x^7+(c+f).x^8
	\end{displaymath}

For the inverse function, for the same reasons as for AOPs, we can just remove the parity bits from the vector obtained after the matrix vector multiplication.

	\begin{displaymath}
		\phi_P^{-1} : b(x) + (a+d).x^6+(b+e).x^7+(c+f).x^8 \rightarrow b(x)
	\end{displaymath}

\subsubsection{The embedding transform}

Like for AOPs, the embedding function for ESPs is direct. 

	\begin{displaymath}
		\phi_E : b(x) \rightarrow b(x) + 0.x^6+0.x^7+0.x^8
	\end{displaymath}

The inverse function consists in computing the remainder modulo $p(x)$, an ESP of degree $r.s$. Thanks to the form of ESP, we can observe that this operation is equivalent to \texttt{xor}ing the last block of $s$ bits to the $r$ blocks of $s$ bits:

	\begin{equation*}
	\phi_E^{-1} :  \begin{array}{c}
	b(x) + g.x^6+h.x^7+i.x^8  \rightarrow \\
	(a+g)+(b+h).x+(c+i).x^2+\\
	(d+g).x^3+(e+h).x^4+(f+i).x^5
	\end{array}
	\end{equation*} 

\section{Implementation}

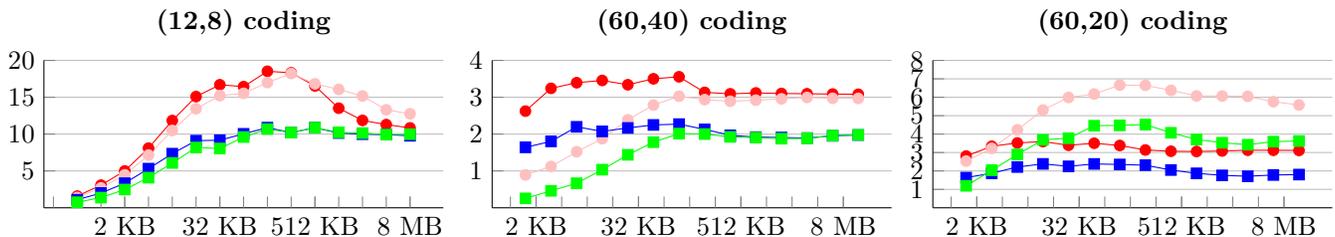
\begin{figure*}[htb!]
	\begin{center}
		\begin{tikzpicture}
		\begin{axis}[
		ymajorgrids,
		width=0.38\textwidth,
		height=0.15\textheight,
		xtick pos=left,
		ytick pos=left,
		legend entries={Pyrit (encoding), Pyrit (decoding), ISA-L (encoding), ISA-L (decoding)},
		legend to name=named,
		xmode = log,
		axis x line*=bottom,
		axis y line*=left,
		ymin=0,
		ymax=20,
		xmin=0,
		ytick={5,10,15,20},
		xtick={128,2048,32768,524288,8388608},
		xticklabels={128 B, 2 KB, 32 KB, 512 KB, 8 MB},
		extra x ticks={128,256,512,1024,2048,4096, 8192, 16384, 32768, 65536, 131072, 262144, 524288, 1048576, 2097152, 4194304, 8388608},
		extra x tick labels={,,,,,,,,,,,,},
		title={\textbf{(12,8) coding}}]
		\addplot+[red, mark=*, mark options={fill=red}] table [x=size, y=t1,, col sep=comma] {encoding_12_8_emb.csv};
		\addplot+[pink, mark=*, mark options={fill=pink}] table [x=size, y=t1,, col sep=comma] {decoding_12_8_emb.csv};
		\addplot+[blue, mark=square*, mark options={fill=blue}] table [x=size, y=t1,, col sep=comma] {encoding_isa_12_8.csv};
		\addplot+[green, solid, mark=square*, mark options={fill=green}] table [x=size, y=t1,, col sep=comma] {decoding_isa_12_8.csv};
		\end{axis}
		\end{tikzpicture}
		\begin{tikzpicture}
		\begin{axis}[
		ymajorgrids,
		width=0.38\textwidth,
		height=0.15\textheight,
		xtick pos=left,
		ytick pos=left,
		legend entries={Pyrit (encoding), Pyrit (decoding), ISA-L (encoding), ISA-L (decoding)},
		legend to name=named,
		xmode = log,
		axis x line*=bottom,
		axis y line*=left,
		ymin=0,
		ymax=4,
		xmin=0,
		ytick={1,2,3,4},
		xtick={128,2048,32768,524288,8388608},
		xticklabels={128 B, 2 KB, 32 KB, 512 KB, 8 MB},
		extra x ticks={128,256,512,1024,2048,4096, 8192, 16384, 32768, 65536, 131072, 262144, 524288, 1048576, 2097152, 4194304, 8388608},
		extra x tick labels={,,,,,,,,,,,,},
		title={\textbf{(60,40) coding}}]
		\addplot+[red, mark=*, mark options={fill=red}] table [x=size, y=t1,, col sep=comma] {encoding_60_40_emb.csv};
		\addplot+[pink, mark=*, mark options={fill=pink}] table [x=size, y=t1,, col sep=comma] {decoding_60_40_emb.csv};
		\addplot+[blue, mark=square*, mark options={fill=blue}] table [x=size, y=t1,, col sep=comma] {encoding_isa_60_40.csv};
		\addplot+[green, solid, mark=square*, mark options={fill=green}] table [x=size, y=t1,, col sep=comma] {decoding_isa_60_40.csv};
		\end{axis}
		\end{tikzpicture}		
		\begin{tikzpicture}
		\begin{axis}[
		ymajorgrids,
		width=0.38\textwidth,
		height=0.15\textheight,
		xtick pos=left,
		ytick pos=left,
		legend columns=4,
		legend entries={Pyrit (encoding), Pyrit (decoding), ISA-L (encoding), ISA-L (decoding)},
		legend to name=named,
		xmode = log,
		axis x line*=bottom,
		axis y line*=left,
		ymin=0,
		ymax=8,
		xmin=0,
		ytick={1,2,3,4,5,6,7,8},
		xtick={128,2048,32768,524288,8388608},
		xticklabels={128 B, 2 KB, 32 KB, 512 KB, 8 MB},
		extra x ticks={128,256,512,1024,2048,4096, 8192, 16384, 32768, 65536, 131072, 262144, 524288, 1048576, 2097152, 4194304, 8388608},
		extra x tick labels={,,,,,,,,,,,,},	
		title={\textbf{(60,20) coding}}]
		\addplot+[red, mark=*, mark options={fill=red}] table [x=size, y=t1,, col sep=comma] {encoding_60_20_par.csv};
		\addplot+[pink, mark=*, mark options={fill=pink}] table [x=size, y=t1,, col sep=comma] {decoding_60_20_par.csv};
		\addplot+[blue, mark=square*, mark options={fill=blue}] table [x=size, y=t1,, col sep=comma] {encoding_isa_60_20.csv};
		\addplot+[green, solid, mark=square*, mark options={fill=green}] table [x=size, y=t1,, col sep=comma] {decoding_isa_60_20.csv};
		\end{axis}
		\end{tikzpicture}
		%
		\caption{Coding speed (GB/s) comparison of MDS erasure codes depending on the block size}
		\label{bench}
	\end{center}
\end{figure*}

We have implemented our method into an MDS erasure code making linear combinations over two finite fields ($\mathbb{F}_{2^4}$ and $\mathbb{F}_{2^6}$) using generalized Cauchy matrices.
The code is written in C, except the coding part which is written  in inline assembly. Indeed, by applying the transforms, we need to generate additional data before doing the \texttt{xor} operations. This data is never used after these xor's. In order to avoid useless memory writes, we compute this additional data into the CPU core by using SIMD registers without writing back the result into the DRAM memory.

For the field $\mathbb{F}_{2^w}$, we use $w$ SIMD registers to load $w$ chunks of a source block, and $w$ other registers to load $w$ chunks of the destination block.

Assume we are working on $\mathbb{F}_{2^4}$ using the SSE SIMD instruction set. We use four 128-bit registers to load 64 bytes of data at a time, which is generally the size of a cache line. Four other registers are used to load the coded data in the same way, and one last register is used to perform the ring transform.

Once the data is loaded into the registers, depending on the finite field and the transform, we use one or three registers to compute the additional data to transform the field elements into ring elements. Then, we do the \texttt{xor} operations before applying the reverse transform to go back from the ring to the field.
As the binary matrices representing the constant elements are only composed by diagonals, we do not need to read the operations to carry out from the memory: for each bit equal to 1 in the constant element, $w$ independent \texttt{xor} operations are unrolled. These binary matrices are built with the "sparse transform" introduced in \ref{sec:sparse}.

Depending on the code parameters, different transforms can be used. Indeed, when $k>=r$, the embedding transform is faster, when $k<r$, the parity transform is faster \cite{isit}. Of course, the corresponding reverse transform must be used.

\section{Performance evaluation}
\label{sec:perf}

In this section, we measure the performances of the encoding and decoding processes. 
For our tests, we used a machine with a 3.40 GHz Intel Core i7-6700 (Skylake architecture) and 16GB of DRAM.
The CPU has 4 cores with hyper threading, 2*32 kB of L1 cache (32 kB of data + 32 kB of instructions) and 256 kB of L2 cache per core, and 8 MB of shared L3 cache.
It supports SSE, AVX and AVX2 instructions sets. 

We defined 3 use cases: a $(12,8)$ over $\mathbb{F}_{2^4}$ with embedding transform, a $(60,40)$ over $\mathbb{F}_{2^6}$ with embedding transform and a $(60,20)$ over $\mathbb{F}_{2^6}$ with parity transform.

We first focused on the excellent Jerasure library \cite{jerasure} because it provides a lot of different methods to perform MDS erasure codes.
In	\cite{pgm:13:sfg}, the authors show that the \textit{split tables} method using SIMD instructions is the fastest implementation for erasure codes based on $\mathbb{F}_{2^w}$ and have the same performances when $w=4$ and $w=8$. 
As we partially did our implementation in assembly code, we fairly compared it with erasure code of the Intel ISA-L library \cite{isal}, another assembly code implementing an erasure code using the \textit{split tables} method with SIMD instructions sets.

As far as we know, ISA-L is the fastest MDS erasure code available because it implements one of the best methods to perform multiplications in a finite field, and it is written in assembly. 

We compared our codec with ISA-L on the 3 different use cases. We also studied the impact of multithreading on the coding speed.

\subsection{Performance analysis using 1 thread}

Figure 2 shows the performances of our codec using  $\mathbb{F}_{2^4}$ or  $\mathbb{F}_{2^6}$ compared to ISA-L. Both codecs are configured to use AVX2 instructions. We encode and decode data by varying the block size for two code parameters: $(12,8)$, $(60,40)$ and $(60,20)$. We repeat the measurements 1000 times for each point.

For the 3 cases, Pyrit is faster than ISA-L for both encoding and decoding. Some reasons are that Pyrit does not use lookup tables like split tables, and needs less instructions to perform the multiplications.
Note that when the total amount of data manipulated ($(k+r)*(symbol size)$) is greater than the L3 cache ($8MB$), the coding speed is constant for both codecs.

\subsection{Performance analysis using multithreading}

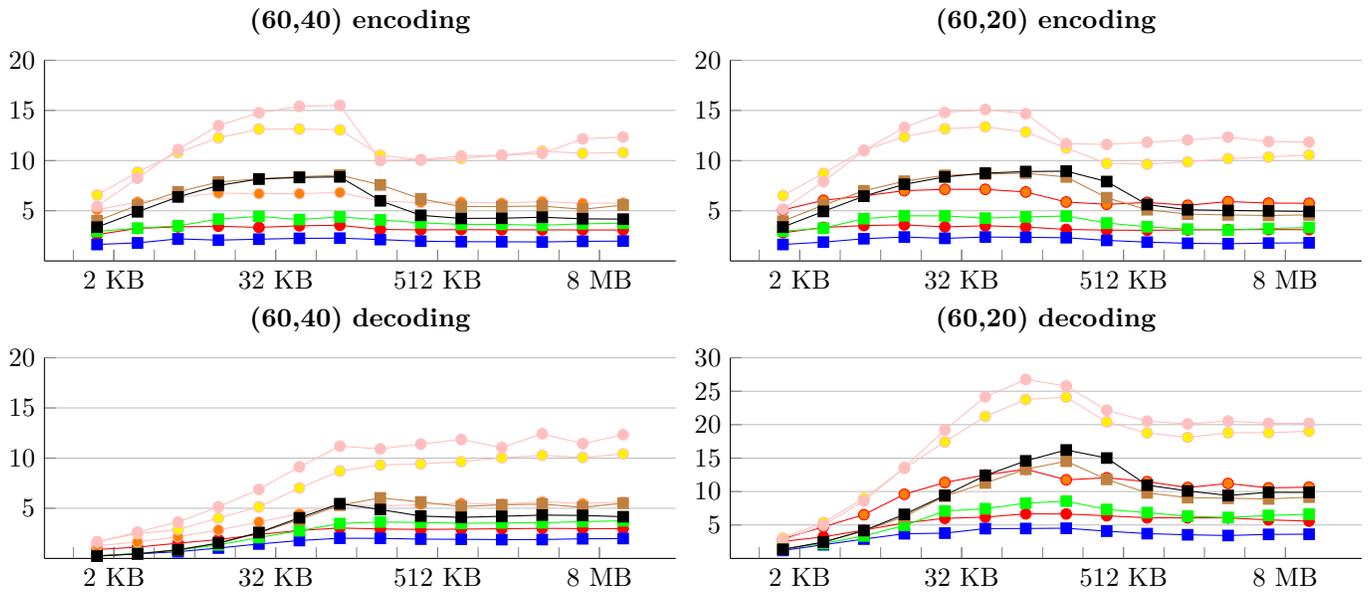
\begin{figure*}
	\begin{center}
		\begin{tikzpicture}
		\begin{axis}[
		ymajorgrids,
		width=0.55\textwidth,
		height=0.18\textheight,
		xtick pos=left,
		ytick pos=left,
		legend entries={Pyrit (1t), Pyrit (2t), Pyrit (4t), Pyrit (8t), ISA-L (1t), ISA-L (2t),ISA-L (4t), ISA-L (8t)},
		legend to name=named,
		xmode = log,
		axis x line*=bottom,
		axis y line*=left,
		ymin=0,
		ymax=20,
		xmin=0,
		ytick={5,10,15,20},
		xtick={128,2048,32768,524288,8388608},
		xticklabels={128 B, 2 KB, 32 KB, 512 KB, 8 MB},
		extra x ticks={128,256,512,1024,2048,4096, 8192, 16384, 32768, 65536, 131072, 262144, 524288, 1048576, 2097152, 4194304, 8388608},
		extra x tick labels={,,,,,,,,,,,,},
		title={\textbf{(60,40) encoding}}]
		\addplot+[red, mark=*, mark options={fill=red}] table [x=size, y=t1,, col sep=comma] {encoding_60_40_emb.csv};
		\addplot+[pink, mark=*, mark options={fill=orange}] table [x=size, y=t2,, col sep=comma] {encoding_60_40_emb.csv};
		\addplot+[pink, mark=*, mark options={fill=yellow}] table [x=size, y=t4,, col sep=comma] {encoding_60_40_emb.csv};
		\addplot+[pink, mark=*, mark options={fill=pink}] table [x=size, y=t8,, col sep=comma] {encoding_60_40_emb.csv};
		\addplot+[blue, mark=square*, mark options={fill=blue}] table [x=size, y=t1,, col sep=comma] {encoding_isa_60_40.csv};
		\addplot+[green, solid, mark=square*, mark options={fill=green}] table [x=size, y=t2,, col sep=comma] {encoding_isa_60_40.csv};
		\addplot+[brown, solid, mark=square*, mark options={fill=brown}] table [x=size, y=t4,, col sep=comma] {encoding_isa_60_40.csv};
		\addplot+[black, solid, mark=square*, mark options={fill=black}] table [x=size, y=t8,, col sep=comma] {encoding_isa_60_40.csv};		
		\end{axis}
		\end{tikzpicture}		
		\begin{tikzpicture}
		\begin{axis}[
		ymajorgrids,
		width=0.55\textwidth,
		height=0.18\textheight,
		xtick pos=left,
		ytick pos=left,
		legend columns=8,
		legend entries={Pyrit (1t), Pyrit (2t), Pyrit (4t), Pyrit (8t), ISA-L (1t), ISA-L (2t),ISA-L (4t), ISA-L (8t)},
		legend to name=named,
		xmode = log,
		axis x line*=bottom,
		axis y line*=left,
		ymin=0,
		ymax=20,
		xmin=0,
		ytick={5,10,15,20},
		xtick={128,2048,32768,524288,8388608},
		xticklabels={128 B, 2 KB, 32 KB, 512 KB, 8 MB},
		extra x ticks={128,256,512,1024,2048,4096, 8192, 16384, 32768, 65536, 131072, 262144, 524288, 1048576, 2097152, 4194304, 8388608},
		extra x tick labels={,,,,,,,,,,,,},	
		title={\textbf{(60,20) encoding}}]
		\addplot+[red, mark=*, mark options={fill=red}] table [x=size, y=t1,, col sep=comma] {encoding_60_20_par.csv};
		\addplot+[red, mark=*, mark options={fill=orange}] table [x=size, y=t2,, col sep=comma] {encoding_60_20_par.csv};
		\addplot+[pink, mark=*, mark options={fill=yellow}] table [x=size, y=t4,, col sep=comma] {encoding_60_20_par.csv};
		\addplot+[pink, mark=*, mark options={fill=pink}] table [x=size, y=t8,, col sep=comma] {encoding_60_20_par.csv};
		\addplot+[blue, mark=square*, mark options={fill=blue}] table [x=size, y=t1,, col sep=comma] {encoding_isa_60_20.csv};
		\addplot+[green, solid, mark=square*, mark options={fill=green}] table [x=size, y=t2,, col sep=comma] {encoding_isa_60_20.csv};
		\addplot+[brown, solid, mark=square*, mark options={fill=brown}] table [x=size, y=t4,, col sep=comma] {encoding_isa_60_20.csv};
		\addplot+[black, solid, mark=square*, mark options={fill=black}] table [x=size, y=t8,, col sep=comma] {encoding_isa_60_20.csv};	
		\end{axis}
		\end{tikzpicture}
		\\
		\begin{tikzpicture}
		\begin{axis}[
		ymajorgrids,
		width=0.55\textwidth,
		height=0.18\textheight,
		xtick pos=left,
		ytick pos=left,
		legend entries={Pyrit (1t), Pyrit (2t), Pyrit (4t), Pyrit (8t), ISA-L (1t), ISA-L (2t),ISA-L (4t), ISA-L (8t)},
		legend to name=named,
		xmode = log,
		axis x line*=bottom,
		axis y line*=left,
		ymin=0,
		ymax=20,
		xmin=0,
		ytick={5,10,15,20},
		xtick={128,2048,32768,524288,8388608},
		xticklabels={128 B, 2 KB, 32 KB, 512 KB, 8 MB},
		extra x ticks={128,256,512,1024,2048,4096, 8192, 16384, 32768, 65536, 131072, 262144, 524288, 1048576, 2097152, 4194304, 8388608},
		extra x tick labels={,,,,,,,,,,,,},
		title={\textbf{(60,40) decoding}}]
		\addplot+[red, mark=*, mark options={fill=red}] table [x=size, y=t1,, col sep=comma] {decoding_60_40_emb.csv};
		\addplot+[pink, mark=*, mark options={fill=orange}] table [x=size, y=t2,, col sep=comma] {decoding_60_40_emb.csv};
		\addplot+[pink, mark=*, mark options={fill=yellow}] table [x=size, y=t4,, col sep=comma] {decoding_60_40_emb.csv};
		\addplot+[pink, mark=*, mark options={fill=pink}] table [x=size, y=t8,, col sep=comma] {decoding_60_40_emb.csv};
		\addplot+[blue, mark=square*, mark options={fill=blue}] table [x=size, y=t1,, col sep=comma] {decoding_isa_60_40.csv};
		\addplot+[green, solid, mark=square*, mark options={fill=green}] table [x=size, y=t2,, col sep=comma] {decoding_isa_60_40.csv};
		\addplot+[brown, solid, mark=square*, mark options={fill=brown}] table [x=size, y=t4,, col sep=comma] {decoding_isa_60_40.csv};
		\addplot+[black, solid, mark=square*, mark options={fill=black}] table [x=size, y=t8,, col sep=comma] {decoding_isa_60_40.csv};		
		\end{axis}
		\end{tikzpicture}		
		\begin{tikzpicture}
		\begin{axis}[
		ymajorgrids,
		width=0.55\textwidth,
		height=0.18\textheight,
		xtick pos=left,
		ytick pos=left,
		legend columns=4,
		legend entries={Pyrit (1t), Pyrit (2t), Pyrit (4t), Pyrit (8t), ISA-L (1t), ISA-L (2t),ISA-L (4t), ISA-L (8t)},
		legend to name=named,
		xmode = log,
		axis x line*=bottom,
		axis y line*=left,
		ymin=0,
		ymax=30,
		xmin=0,
		ytick={5,10,15,20,25,30},
		xtick={128,2048,32768,524288,8388608},
		xticklabels={128 B, 2 KB, 32 KB, 512 KB, 8 MB},
		extra x ticks={128,256,512,1024,2048,4096, 8192, 16384, 32768, 65536, 131072, 262144, 524288, 1048576, 2097152, 4194304, 8388608},
		extra x tick labels={,,,,,,,,,,,,},	
		title={\textbf{(60,20) decoding}}]
		\addplot+[red, mark=*, mark options={fill=red}] table [x=size, y=t1,, col sep=comma] {decoding_60_20_par.csv};
		\addplot+[red, mark=*, mark options={fill=orange}] table [x=size, y=t2,, col sep=comma] {decoding_60_20_par.csv};
		\addplot+[pink, mark=*, mark options={fill=yellow}] table [x=size, y=t4,, col sep=comma] {decoding_60_20_par.csv};
		\addplot+[pink, mark=*, mark options={fill=pink}] table [x=size, y=t8,, col sep=comma] {decoding_60_20_par.csv};
		\addplot+[blue, mark=square*, mark options={fill=blue}] table [x=size, y=t1,, col sep=comma] {decoding_isa_60_20.csv};
		\addplot+[green, solid, mark=square*, mark options={fill=green}] table [x=size, y=t2,, col sep=comma] {decoding_isa_60_20.csv};
		\addplot+[brown, solid, mark=square*, mark options={fill=brown}] table [x=size, y=t4,, col sep=comma] {decoding_isa_60_20.csv};
		\addplot+[black, solid, mark=square*, mark options={fill=black}] table [x=size, y=t8,, col sep=comma] {decoding_isa_60_20.csv};	
		\end{axis}
		\end{tikzpicture}		
		\label{mt}
		\caption{Coding speed (GB/s) comparison depending on the block size and the number of threads}
		\label{bench}
	\end{center}
\end{figure*}

We modified both ISA-L and Pyrit to support multithreading coding. To do that, we split the data to encode or decode, and send a different part to all the threads, synchronized with pthread barriers. That means that each thread is writing and reading a different part of the symbols to code. Figure 3 shows the results for 1,2,4 and 8 threads using ISA-L and Pyrit (over $\mathbb{F}_{2^4}$).

Both codecs have increased performances using multiple threads, even if Pyrit is still faster than ISA-L.
But, when the number of threads is greater than the number of cores, for $8$ threads on $4$ cores, the performance of the ISA-L codec decreases whereas this is not the case for Pyrit, even if the gain is low.

\subsection{Results}

For any coding parameters, the Pyrit method performs much faster than ISA-L. Even when the data does not fit into the CPU caches, our method drastically improves the performances.
Making linear combinations over $\mathbb{F}_{2^4}$ or $\mathbb{F}_{2^6}$ using Pyrit gives the same performances for the encoding process. This can be explained since we use a generalized Cauchy matrix which is sparse but generic matrix. For the decoding process, as the inverse matrix depends on the loss pattern, we have a slightly dense matrix using $\mathbb{F}_{2^6}$, so working in $\mathbb{F}_{2^4}$ is faster.
The only constraint is the number of symbols which is respectively $16$ and $64$ for $\mathbb{F}_{2^4}$ and $\mathbb{F}_{2^6}$. With ISA-L, it is possible to have $256$ symbols. Nevertheless, when $n<=64$, Pyrit is the fastest method.

\section{Conclusion}
\label{sec:ccl}

In this paper, we have presented a new method to accelerate both coding and decoding processes of erasure codes. By using transforms between a finite field and a polynomial ring, sparse generator matrices can be obtained. This allows to significantly reduce the complexity of matrix vector multiplication. 
We presented two fast erasure codes implementations using $\mathbb{F}_{2^4}$ and $\mathbb{F}_{2^6}$ and evaluated the performances for several use cases. For each cases, Pyrit has better performances. The performance analysis was done for MDS erasure codes, but Pyrit can be used by every code using matrix vector multiplication over a finite field.

\bibliographystyle{IEEEtran}
	\bibliography{biblio}
\end{document}